\documentclass{ws-procs9x6}

\begin{document}
\title{WORLDLINE APPROACH TO QFT\\ ON MANIFOLDS WITH BOUNDARY}
\author{F. BASTIANELLI and \underline{O. CORRADINI}}
\address{Dipartimento di Fisica, Universit\`a di Bologna
and INFN, Sezione di Bologna\\
Via Irnerio 46, I-40126 Bologna, Italy\\
E-mail: lastname@bo.infn.it}

\author{P. A. G. PISANI}
\address{IFLP (CONICET), Departamento de F\'isica
de la Universidad 
Nacional de La Plata\\ c.c. 67, 1900 La Plata,
Argentina\\
E-mail: pisani@obelix.fisica.unlp.edu.ar} 

\author{C. SCHUBERT}
\address{Instituto de F\'{\i}sica y Matem\'aticas,
Universidad Michoacana de San Nicol\'as de Hidalgo, 
Edificio C-3, Apdo. Postal 2-82,
C.P. 58040, Morelia, Michoac\'an, M\'exico\\
E-mail: schubert@ifm.umich.mx}
\begin{center}{\em To appear in the proceedings of conference QFEXT09}
\end{center}

\begin{abstract}
{\bf Abstract:} We use the image charge method to compute the trace of
the heat kernel for a scalar 
field on a flat manifold with boundary, representing the trace 
by means of a worldline path integral and obtain useful non-iterative master formulae for $n$
insertions of the scalar potential. We discuss possible extensions of the method.  
\end{abstract}


\bodymatter

%
\section{Worldline formalism on manifolds with boundary}
The worldline formalism (see Ref.~\refcite{Schubert:2001he} for a
review) is an alternative method to compute effective actions,
amplitudes and anomalies in quantum field theory. For example, the
one-loop effective action can be written as a 
``trace log'' of a differential operator which be exponentiated
using a Schwinger proper time integral and the trace can be written in
terms of a quantum mechanical path integral. For the simplest case of a
real massless scalar field with self-interaction $U(\phi)$, propagating
in a flat boundaryless space, the one-loop
effective action formally reads
$
\Gamma[\varphi] = -\frac12 \int_0^\infty \frac{dT}{T} {\rm Tr}\ e^{-T H} 
$
where $H=-\Box +U''(\phi)$ and the partition trace reads
\begin{equation}
  {\rm Tr}\ e^{-TH} = \int_{PBC} Dx\ \exp\Biggl[-\int_0^1d\tau \left(
   \frac1{4T} \dot x^2 +T\, V(x(\tau))\right) \Biggr]
\label{trace:PI}  
\end{equation}
where $V(x) \equiv U''(\varphi(x))$ and the path integral is over the
space of all closed paths on the unit circle and the
Teichmuller parameter $T$ is the proper length of the circle. The
short-time expansion of the operator ${\rm Tr}\ e^{-TH}$, known as heat
kernel expansion, takes the form
\begin{equation}
{\rm Tr}\ e^{-TH} = \int d^D x\ K(T;x,x)  =\frac1{(4\pi T)^{D/2}}
\sum_{n=0}^\infty a_n\ T^{n}  
\label{trace}
\end{equation}
and the integrated heat kernel coefficients $a_n$ can be straightforwardly
obtained as a short-time expansion of the path
integral~(\ref{trace:PI}) by Wick contracting the Taylor
expansion of the potential~\cite{Fliegner:1993wh}.

The worldline path integral approach to QFT on manifolds with
boundary has been carried out using Monte Carlo
simulations~\cite{Gies:2003cv} and many interesting results are obtained with this
method~\cite{Gies:QFexT09}. However, a serious difficulty one has to
face with such a method concerns boundary 
conditions different than Dirichlet and an alternative
method that might help overcoming this difficulty would be quite
welcome. 

In a manifold with boundary a heat kernel expansion for the
trace, cfr. Eq.~(\ref{trace}), still holds but the sum involves half-integer powers and the
coefficients include boundary contributions as well as bulk
contributions~\cite{McKean:1967xf}. We have
developed~\cite{Bastianelli:2006hq}
a method that generalizes analytic 
worldline techniques to flat manifolds with boundary $M={\mathbb
  R}_+\times {\mathbb R}^{D-1}$ using the image charge method to map
the path integral on a half space to the combination of two path
integrals on the whole space
\begin{eqnarray}
{\rm Tr}_M\ e^{-TH} = \int_M d^D x\ K(T;x,x) & \mp &\int_M d^D
x\ K(T;\tilde x,x)
\\ && x=(y,{\vec z})\,,\quad y\in{\mathbb
  R}_+\,,\ {\vec z} \in {\mathbb R}^{D-1} 
\nonumber
\end{eqnarray}        
where $\tilde x=(-y,{\vec z})$ is the image
charge of $x$ and  the upper (lower) sign corresponds to Dirichlet (Neumann) boundary
conditions 
(in Ref.~\refcite{Bastianelli:2007jr} an extension of the
method to Robin
boundary conditions was considered). 
The above kernels are whole space kernels computed
with an evenly extended potential 
\begin{equation}
V(x)\ \to \ \tilde V(x) = \theta(y) V(x) +\theta(-y) V(\tilde x) = V_+(x)
+\epsilon(y) V_-(x)~.
\end{equation}
The reflection property of the potential also allows to extend the
overall Riemannian integral to the whole space, so that the above two
contributions can be written as
\begin{eqnarray}
K^{dir}(T) &\equiv& \int_M d^D x\ K(T;x,x) =\frac12 \int d^D
x\ K(T;x,x)\nonumber\\
 &=& \frac12 \int_{PBC} Dx\ \exp\Biggl[-\int_0^1d\tau \left(
   \frac1{4T} \dot x^2 +T\, \tilde V(x(\tau))\right) \Biggr]\\   
K_{\partial M}^{ind}(T) &\equiv& \int_M d^D x\ K(T;\tilde x,x) =\frac12 \int d^D
x\ K(T;\tilde x,x)\nonumber\\
 &=& \frac12 \int_{(A)PBC} Dx\ \exp\Biggl[-\int_0^1d\tau \left(
   \frac1{4T} \dot x^2 +T\, \tilde V(x(\tau))\right) \Biggr]   
\label{K-ind}
\end{eqnarray}
and will be referred to as the ``direct contribution'' and the ``indirect
contribution'' respectively. Above, the suffix ${}_{(A)PBC}$ indicates that
the coordinate $y(\tau)$ satisfies (anti)-periodic boundary 
conditions, whereas coordinates ${\vec z}(\tau)$ satisfy periodic boundary conditions.  
The potential $\tilde V$ includes a distribution
$\epsilon(y)$ and the naive application of
the Wick theorem  results
nontrivial~\cite{Bastianelli:2006hq,Bastianelli:2007jr}. However,
we~\cite{Bastianelli:2008vh} demonstrated that (i) upon Fourier representing
the sign function 
$\epsilon(y) = \int \frac{dp}{\pi p} \sin(p y) = \int_{ev} \frac{dp}{i
  \pi p}\ e^{i p y}  
$
and (ii) upon carefully separating out bulk contributions from boundary
contributions, one can safely use it:
two new coefficients for the half-space, $a_4$ and $a_{9/2}$, were computed.

\subsection{Indirect contribution to the heat kernel trace}
For this contribution the coordinate $y(\tau)$ is antiperiodic 
and therefore its kinetic action has no zero mode. Hence,
$\int_{ABC}Dy\ e^{-\frac1{4T}\int_0^1d\tau \dot y^2} =\frac12$
 and we can safely Taylor expand the potential about
the boundary $(0,{\vec z})$
\begin{equation}
\tilde V(y(\tau),{\vec z} +{\vec z}(\tau)) = e^{{\vec
    z}(\tau)\cdot{\vec \partial}}\Biggl[e^{y(\tau)\partial_y}
  V_+(0,{\vec z})
+\int_{ev}\frac{dp}{i\pi p} e^{y(\tau)D^0(p)}V_-(0,{\vec z})
\Biggr]
\end{equation}  
with $D^0(p) = \partial_y +i p$. Inserting the latter
into~(\ref{K-ind}) one obtains 
\begin{eqnarray}
K_{\partial M}^{ind}(T) &=& \frac1{4(4 \pi
  T)^{\frac{D-1}{2}}}\sum_{n=0}^\infty \frac{(-T)^n}{n!}\int_0^1
d\tau_1\cdots\int_0^1 d\tau_n \int_{\partial M}d^{D-1} z\nonumber\\
&\times & \exp\Biggl[-\frac T2 \sum_{i,j=1}^n \left(
  G_P(\tau_i,\tau_j) \vec\partial_i\cdot \vec\partial_j
  +G_A(\tau_i,\tau_j) D^0_i(p) D^0_j(p)\right)\Biggr]\nonumber\\
&\times &\prod_{k=1}^n\Biggl[ V_+^{(k)}(0,\vec z) +
  \int_{ev}\frac{dp_k}{i \pi p_k}V_-^{(k)}(0,\vec z)\Biggr]    
\label{K-ind-gen}
\end{eqnarray}
and the suffix $i$ on the derivative means that it acts on the term of potential
labelled accordingly. The ABC propagator appearing above is given by 
$G_A(\tau,\sigma)=|\tau -\sigma|-\frac12$,
whereas the expression for the PBC propagator depends on the
prescription one adopts for factoring out 
the zero mode $\vec z$. For example, in the "String Inspired'' method
we have $G_P(\tau,\sigma) = |\tau
-\sigma|-(\tau -\sigma)^2$, and using (worldline)
DBC we instead have $G_{P}(\tau,\sigma) = |\tau
-\sigma|+\frac12(1-2\tau)(1-2\sigma)-\frac12$.  The two methods yield
different unintegrated heat kernel 
expansions and their difference resides on total derivative
terms. However, since here these terms are boundary
total derivatives their integrals vanish and the integrated
expression~(\ref{K-ind-gen}) is
scheme-independent. 
Scheme-independence will be slightly more subtle
for the direct contribution that we describe next.       

\subsection{Direct contribution to the heat kernel trace}
Here all the coordinates have periodic boundary
conditions and we Taylor expand the potential insertions about
the zero modes $(y,\vec z)$ and get 
\begin{eqnarray}
\hspace{-15pt}
K^{\rm dir}(T)
&=&
\frac1{2(4\pi T)^{{\frac D2}}}
\sum_{n=0}^{\infty}
{\frac{(-T)^n}{n!}}
\int_0^1d\tau_1 \cdots \int_0^1d\tau_n\int_{-\infty}^\infty dy\int_{\partial M} d^{D-1}z
\nonumber\\
&& 
\hspace{-25pt}
\times
{\rm exp}\biggl[-\frac{T}{2}
\sum_{i,j=1}^n G_{P}(\tau_i,\tau_j)\partial_i\cdot\partial_j
\biggr]
\prod_{k=1}^n
\, \Bigl\lbrack V^{(k)}_+ (x) + (\epsilon (y)V_-(x))^{(k)} \Bigr\rbrack
\label{Kdirmaster}
\end{eqnarray}
where the notation is meant to convey that derivatives may act on
$\epsilon(y)$ as well as on $V_\pm$. When derivatives act on
$\epsilon(y)$, $\delta$ functions or derivatives thereof are generated
giving rise to boundary terms. Contributions where no derivatives act
on $\epsilon(y)$ are bulk terms. Namely 
\begin{eqnarray}
\hspace{-15pt}
K_M^{\rm dir}(T)
&=&
\frac1{2(4\pi T)^{{\frac D2}}}
\sum_{n=0}^{\infty}
{\frac{(-T)^n}{n!}}
\int_0^1d\tau_1 \cdots \int_0^1d\tau_n\int_{-\infty}^\infty dy\int_{\partial M} d^{D-1}z
\nonumber\\
&& 
\hspace{-25pt}
\times
{\rm exp}\biggl[-\frac{T}{2}
\sum_{i,j=1}^n G_{D,S}(\tau_i,\tau_j)\partial_i\cdot\partial_j
\biggr]
\prod_{k=1}^n
\, \Bigl\lbrack V^{(k)}_+ (x) + \epsilon (y)V^{(k)}_-(x) \Bigr\rbrack~.
\label{KdirMmaster}
\end{eqnarray}
Subtracting~(\ref{KdirMmaster}) from~(\ref{Kdirmaster}) yields the
boundary terms associated to the direct contribution, namely      
\begin{eqnarray}
K_{\partial M}^{\rm dir}(T)
&=& 
\frac1{2
(4\pi T)^{\frac{D}{2}}}
\sum_{n=0}^{\infty}
\frac{(-T)^n}{n!}
\int_0^1d\tau_1 \cdots \int_0^1d\tau_n
\int_{-\infty}^{\infty}dy
\int_{\partial M} d^{D-1}z
\nonumber\\
&& \hspace{-10pt}
\times \int_0^1dw \, \frac{\partial}{\partial w}
\,{\rm exp}\biggl[-\frac{T}{2}
\sum_{i,j=1}^n G_P(\tau_i,\tau_j)D_i(wp)\cdot D_j(wp)
\biggr]
\nonumber\\
&&\hspace{-10pt}\times
\prod_{k=1}^n
\, \Bigl\lbrack V^{(k)}_+(x)+ \int_{ev}\frac{dp_k}{i\pi p_k}
\, e^{ip_k y}V_-^{(k)}(x) \Bigr\rbrack
\label{KdirboundarymasterD}
\end{eqnarray}
where $D(wp)=(D^0(wp),\vec \partial)$  and
the total derivative on $w$ takes care of the aforementioned
subtraction. The evaluation of~(\ref{KdirboundarymasterD}) is done as
follows: 
\begin{romanlist}[4]
\item Taylor expand potentials about the boundary: it is safe as by
  construction all terms in~(\ref{KdirboundarymasterD}) are boundary
  terms;
\item integrate over $y$: it yields a $\delta$ function involving
  various $p$'s and $\partial$'s;
\item $w$ derivative cancels one (spurious) pole;
\item integrate over all $p$'s, then over $w$ and finally over
  $\tau_i$.
\end{romanlist}     
Expressions~(\ref{KdirMmaster}) and~(\ref{KdirboundarymasterD})
separately are scheme-dependent but the scheme-dependent terms cancel out in the
sum~(\ref{Kdirmaster}). 
In fact as mentioned above scheme-dependence
of the bulk part is encoded in a set of total derivative terms that
upon integration yield boundary terms.

%
\section{Outlook}
We discussed a path integral method to compute the heat kernel trace for
a self-interacting scalar field on a flat manifold with boundary. A
natural generalization is the inclusion of interaction with external
fields. This is clearly feasible by evenly extending to the whole
space the coupling $A_M(y,\vec z) x^M$ or $h_{MN}(y,\vec z) x^M x^N$ for the spin-one and
spin-two case. Another possible generalization involves the inclusion
of particles with spin in the loop, e.g. by representing the effective
action in terms of a spinning particle action on the circle.   

%
\section*{Acknowledgments}
The work of F.B. and O.C. was
partly supported by the Italian MIUR-PRIN contract 20075ATT78. 
 The work of P.P. was partly supported by PIP 6160, UNLP proj. 11/X381 and DAAD.

%
\bibliographystyle{ws-procs9x6}
\bibliography{OC_QFexT}

\end{document}